\documentclass[reprint, aps,prb,lenghtcheck]{revtex4-1}
\usepackage{amsfonts}
\usepackage{graphicx}
\usepackage{pdfpages} % include the figure pages
\usepackage{color}
\usepackage{amsmath}
\newcommand{\K}[1]{$|#1\rangle$}
\newcommand{\La}{{\cal L}}

\newcommand{\bea}{\begin{eqnarray}}
\newcommand{\eea}{\end{eqnarray}}

\graphicspath{{Figures/}}

\begin{document}

\title{Measurement crosstalk between two phase qubits coupled by a coplanar waveguide}
\author{Fabio Altomare$^1$}
\author{Katarina Cicak$^1$}
\author{Mika A.  Sillanp\"a\"a$^2$}
\author{Michael S. Allman$^{1,3}$}
\author{Adam J. Sirois$^{1,3}$}
\author{Dale Li$^1$}
\author{Jae I. Park$^1$}
\author{Joshua~A. Strong$^{1,3}$}
\author{John D. Teufel$^{1}$}
\author{Jed D. Whittaker$^{1,3}$}
\author{Raymond W. Simmonds$^1$}
\affiliation{$^1$National Institute of Standards and Technology, 325
Broadway, Boulder CO 80305, USA\\
$^2$Helsinki University of Technology, Espoo P.O. Box
2200 FIN-02015 HUT Finland\\
$^3$University of Colorado, 2000 Colorado Ave, Boulder, CO 80309-0390, USA}
\begin{abstract}

We analyze the measurement crosstalk between two flux-biased phase
qubits coupled by a resonant coplanar waveguide cavity. After the
first qubit is measured, the superconducting phase can undergo
damped oscillations resulting in an a.c. voltage that produces a
frequency chirped noise signal whose frequency crosses that of the
cavity. We show experimentally that the coplanar waveguide cavity
acts as a bandpass filter that can significantly reduce the
crosstalk signal seen by the second qubit when its frequency is
far from the cavity's resonant frequency. We present a simple
classical description of the qubit behavior that agrees well with
the experimental data. These results suggest that measurement
crosstalk between superconducting phase qubits can be reduced by
use of  linear or possibly nonlinear resonant cavities as coupling
elements.
\end{abstract}

\maketitle

In recent years, much effort has been spent fabricating
superconducting circuits with embedded Josephson junctions (JJs)
as a promising platform for developing a quantum computer. In
particular, superconducting qubits, broadly classified as
charge, flux, and phase\cite{[For a recent review see ][ and references therein]{ClarkeN2008}}, have
recently achieved coherence time longer than
7~$\mu$s\cite{WallraffPRL2005}, and single shot visibility close
to 90\%\cite{LupascuPRL2006,*MartinisQIP2009}. Various schemes
have been devised to couple several qubits in a more complex
circuit: coupling through JJs\cite{GrajcarPRL2006}, inductive
coupling\cite{NiskanenSci2007,*NiskanenPRB2006}, capacitive
coupling\cite{McDermottSci2005} and coupling through a resonant
coplanar waveguide\cite{MajerNat2007,Sillanpaeae2007,AnsmannN2009}
(CPW) cavity have all been achieved. The quantum mechanical nature
of CPW cavities has also been demonstrated by generating arbitrary
Fock states through the use of a coupled phase
qubit\cite{HofheinzN2009}. Additionally, a protocol  for the
preparation of arbitrary entangled states of two phase qubits and
a CPW cavity has been developed\cite{AltomareX2009}.

Here, we will focus on two flux-biased phase
qubits\cite{SimmondsPRL2004} coupled through a CPW
cavity\cite{Sillanpaeae2007,AltomareX2009,AnsmannN2009} and will
show how the CPW cavity plays a crucial role in the reduction of
measurement crosstalk\cite{Sillanpaeae2007,AnsmannN2009}.

\begin{figure} % Figure 1
\includegraphics[width=\columnwidth, trim=0 1cm 0 0]{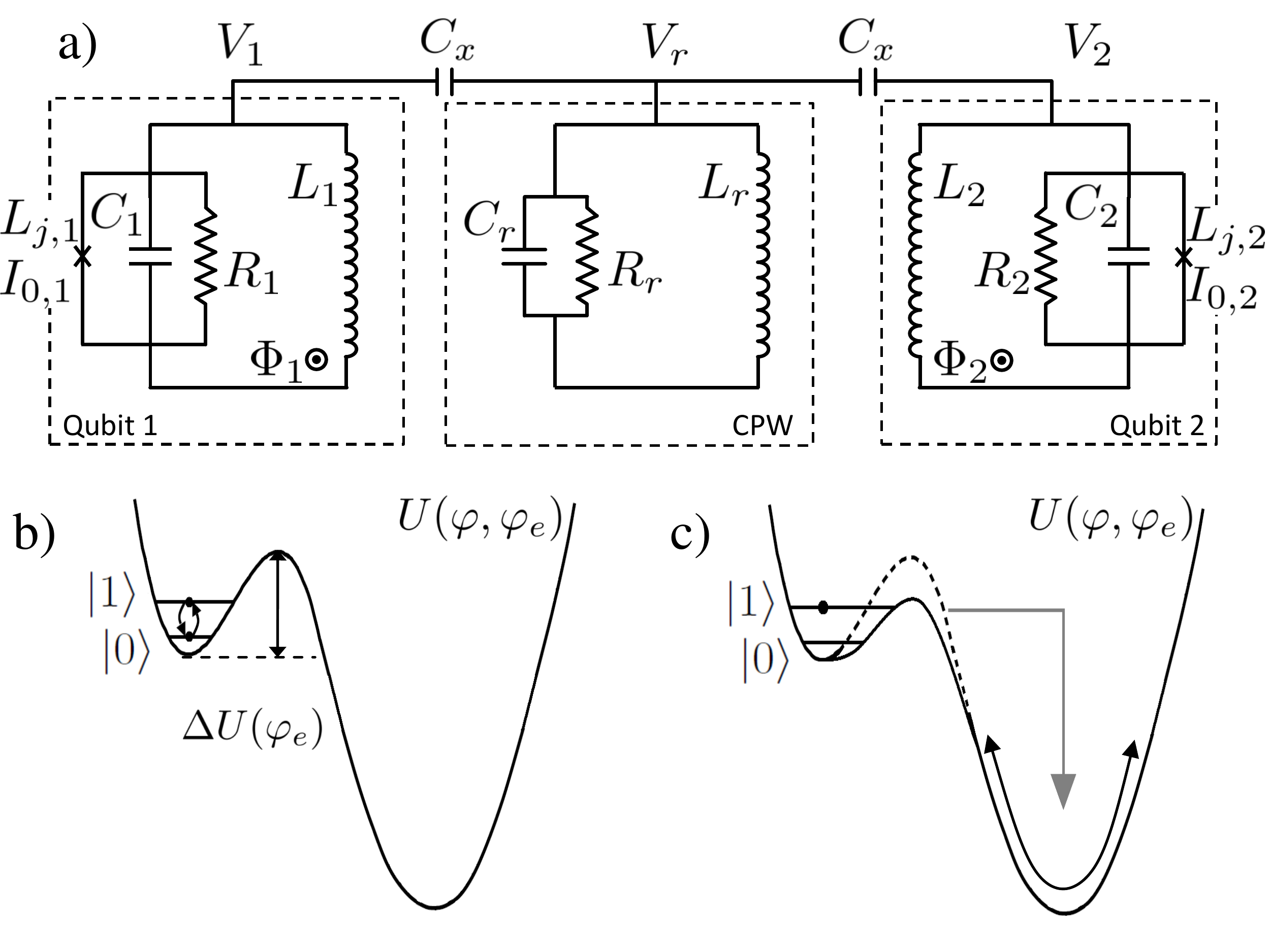}
\caption{(a) Equivalent electrical circuit for two flux-biased
phase qubits coupled to a CPW cavity (modelled as a lumped element
harmonic oscillator). $C_i$ is the total $i-$qubit (or CPW cavity)
capacitance, $L_i$ the geometrical inductance, $L_{j,i}$ the
Josephson inductance of the JJ, $R_i$ models the dissipation in
the system. (b) $U(\varphi,\varphi_e)$ is the potential energy of
the phase qubit as function of superconducting phase difference
$\varphi$ across the JJ and the   dimensionless external flux bias
$\varphi _e=\Phi 2\pi/\Phi_0$. $\Delta U(\varphi _e)$ is the
difference between the local potential maximum and the local
potential minimum in the left well at the flux bias $\varphi _e$.
(c) During the MP, the potential barrier $\Delta U(\varphi _e)$
between the two wells is lowered for a few nanoseconds allowing
the \K{1} state to tunnel into the right well where it will
(classically) oscillate and lose energy due to the dissipation.
\label{fig:QBpotential}}
\end{figure}

Measurement crosstalk in coupled flux-biased phase
qubits\cite{McDermottSci2005,kofmanprb2007} results from their
unique formation in a metastable well of a double-well potential
and the measurement scheme used for determining the qubit state.
The schematic of a typical phase qubit\footnote{In the remainder,
we will refer to the flux-biased phase qubit simply as a phase
qubit} circuit is shown in Fig.~\ref{fig:QBpotential}(a). The phase qubit is essentially a resonant LC circuit in parallel with a Josephson inductance ($L_J=h/2e/(2\pi I(\Phi))=\Phi_0/(2\pi
I_0\cos(2\pi\Phi/\Phi_0))$), where $\Phi_0$ is th flux quantum, $\Phi$ is the flux applied to the circuit loop and $I_0$ the JJ critical current. 
The potential energy as a function of the superconducting phase difference ($\varphi$) across the JJ is presented in
Fig.~\ref{fig:QBpotential}(b) for a particular flux bias. With a
relatively strong anharmonicity, the quantized energy levels in
the left well can be individually addressed with transition
frequencies between the lowest quantized state (\K{0}) and the
first excited state (\K{1}) in the microwave region. The
occupation probability of the qubit's first excited state is measured by applying  a fast flux pulse or measure pulse (MP)
that tilts the well for a few nanoseconds\cite{CooperPRL2004} so
that only the \K{1} state can tunnel out to the right well, as
shown in Fig.~\ref{fig:QBpotential}(c). Because the phase qubit is
formed in a metastable region of the potential, its 'ground state'
energy is naturally higher than the global ground state of the
system. Upon tunneling, this additional energy is released so that
the phase of the qubit (classically) undergoes large oscillations
in the deeper right well. Following the MP the flux is adjusted to
form a symmetric double-well potential. After tens of
microseconds, when the system has relaxed due to dissipation, a DC
SQUID detects the flux in the qubit loop, allowing us to
discriminate between the two circulating current states where the
phase either relaxed in the left well or the right well. These
correspond to the qubit states \K{0} if the qubit did not tunnel
and \K{1} if the qubit did tunnel.

The oscillations of the qubit phase in the right well produce an
oscillating voltage across the JJ\cite{JosephsonRMP1974} with a
relatively large size, representing roughly hundreds of microwave
quanta. This voltage signal can excite any devices coupled to the
qubit when their resonant frequency matches that of the
oscillation. Because the right well potential is weakly
anharmonic, as the amplitude decreases due to dissipation the
frequency rises, producing a chirp crosstalk signal spanning over
10 GHz. With direct capacitive coupling between two phase
qubits\cite{McDermottSci2005}, this process results in the second
qubit being excited over its metastable barrier whenever the first
qubit is measured in the \K{1} state. Due to the nonlinear
dynamics of the system, there is a finite amount of time required (several
nanoseconds) for the second qubit to be excited after the
tunneling of the first qubit. It was found that the loss of qubit
information due to measurement crosstalk could be avoided by
measuring the two qubits simultaneously\cite{McDermottSci2005}
(within $\sim 2$~ns). A classical description of the qubit
dynamics was found to sufficiently model the observed measurement
crosstalk behavior\cite{McDermottSci2005,kofmanprb2007}.
Unfortunately, the excess energy released by a metastable phase
qubit during a tunnelling event is unavoidable, so that
measurement crosstalk can be a serious problem for systems with
many coupled qubits.

The situation is quite different if two phase qubits are coupled
through a CPW cavity. Here we show both experimentally and using
the classical description of Ref.~\onlinecite{kofmanprb2007} that
the CPW cavity acts as a bandpass filter, and the second device is
excited only if it is in resonance with the CPW cavity. This
suggests a simple and effective way to reduce the measurement
crosstalk between coupled devices and has recently been
implemented\cite{AnsmannN2009}.

The device we use has already been described
elsewhere~\cite{AltomareX2009}, and consists of two flux-biased
phase qubits capacitively coupled through a 7~mm open-ended
coplanar waveguide whose half-wave resonant mode frequency is
$\omega_r/2\pi\approx 8.9$~GHz. All the measurements are performed
at a temperature of 35 mK in a dilution refrigerator. The qubits
are controlled through heavily filtered (low pass) flux bias
lines, while the microwave lines  are attenuated at various stages
in the cryostat. Figure~\ref{fig:QBpotential}(a) shows the
equivalent electrical circuit for the device. The dissipation in
the circuit is modelled using the RSJ
model\cite{barone1982physics,likharev1986dynamics,vanduzer1981principles}.
The CPW cavity, with characteristic impedance $Z_r\approx 50
~\Omega$ close to resonance, is equivalent to a lumped element
resonator with $L_r= 2Z_r/ \pi\omega_r\approx 570~pH$,
$C_r=\pi/2\omega_rZ_r\approx 0.56~pF$. The lifetime of an
excitation in the CPW cavity\cite{SimmondsQIP2009} is
$T^{r}_1\approx~1 \mu$s,  which yields
$R_r=T^{r}_1/C_r=1.8~M\Omega$. The qubit's parameters are:
$L_1=690$\ pH, $C_1=0.7$~pF, $I_1=0.8~ \mu$A, $T^{1}_1=170$~ns,
$R_1=240~k\Omega$, and $L_2=690$~pH, $C_2=0.7$~pF,
$I_2=0.95$~$\mu$A, $T^{2}_{1}=70$~ns, $R_2=100~k\Omega$, with
$C_{x}=6.2$~fF. As discussed in Ref.~\onlinecite{AltomareX2009},
the resonant frequency of both qubits exhibits an avoided crossing
at the CPW cavity frequency ($\approx 8.9$~GHz). For the first
qubit this happens at a flux $\overline{\Phi}_1 = 0.82\Phi_{c1}$,
and for the second at a flux $\overline{\Phi}_2=0.842\Phi_{c2}$.
For each qubit, $\Phi_{ci}$ is the critical flux at which the left
well of Fig.~\ref{fig:QBpotential}(b) disappears.

\begin{figure} % Figure 3
\includegraphics[trim=0 2cm 0 2cm,width=\columnwidth]
{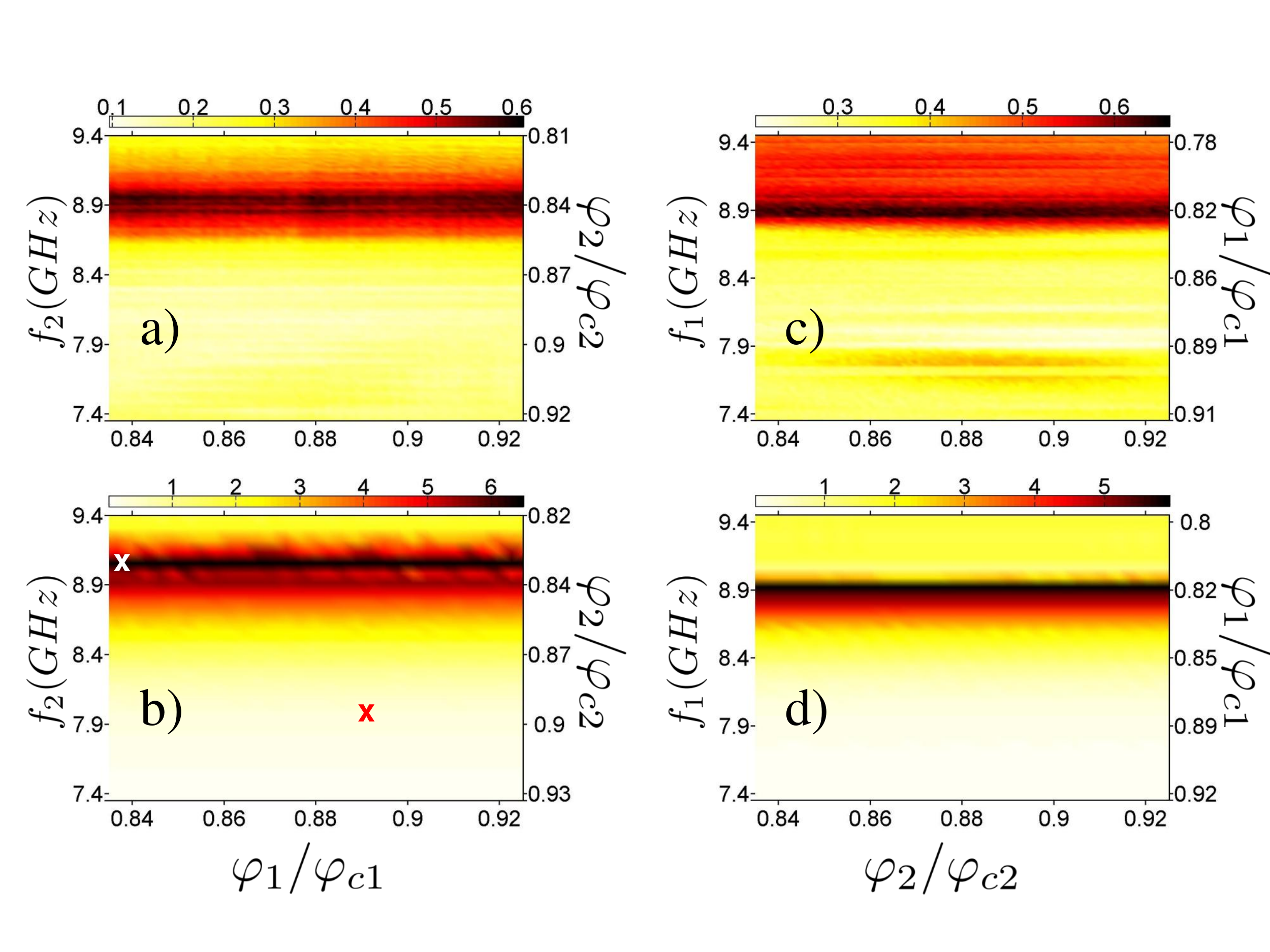} \caption{(Color) Measurement crosstalk: (a)
Experimental tunneling probability for qubit 2, after qubit 1 has
already tunneled as function of the (dimensionless) flux applied
to the qubits. The left ordinate displays the resonant frequency
as measured from the qubit spectroscopy. The right ordinate
displays the ratio between the applied flux and the critical flux
for qubit 2. (b) Simulation: ratio between the maximum energy
acquired by the second qubit and the resonant frequency in the
left well  ($N_l$) as a function of the flux applied to the
qubits. The left ordinate displays the oscillation frequency as
determined from the Fast Fourier Transform of the energy of qubit
2. The right ordinate displays the ratio between the applied flux
and the critical flux for qubit 1. Temporal traces corresponding
to the two \textbf{x}'s are displayed in
Fig.~\ref{fig:singlelinecut}. (c-d) Same as (a-b) after reversing
the roles of the two qubits. \label{fig:experiment}}
\end{figure}

For our experiment, we initially
determine\cite{McDermottSci2005,Sillanpaeae2007,AltomareX2009} the
optimal 'simultaneous' timing between the two MPs that takes into
account the different cabling and instrumental delays from the
room-temperature equipment to the cold devices. Then, as a function
of the flux applied to the two qubits, we measure the tunneling
probability for the second (first) qubit after we purposely induce
a tunneling event in the first (second) qubit. The results are
shown in Fig.~\ref{fig:experiment}(a,c). The probability of
finding the second (first) qubit in the excited state as a result
of measurement crosstalk is significant only in a region around
$\varphi_2/\varphi_{c2}=\Phi_2/\Phi_{c2}\approx
\overline{\Phi}_2/\Phi_{c2}\sim 0.842$
($\varphi_1/\varphi_{c1}=\Phi_1/\Phi_{c1}\approx
\overline{\Phi}_1/\Phi_{c1}\sim 0.82$) where the resonant
frequency of the second (first) qubit is close to the CPW cavity
frequency.
\\

To provide a qualitative description of these results,  we write
the Lagrangian\cite{kofmanprb2007} for the two qubits coupled
through a CPW cavity (Fig.~\ref{fig:QBpotential} (a)) as
\begin{equation}
\begin{split}
  \La = {} &  \frac{1}{2}C_1V_1^2 +\frac{1}{2}C_2V_2^2+\frac{1}{2}C_rV_r^2 + \frac{1}{2}C_{x}(V_1-V_r)^2+\\
  {} & \frac{1}{2}C_{x}(V_r-V_2)^2
       -U_1(\phi) -U_2(\phi) - \frac{\Phi_r^2}{2L_r}
\end{split}
 \end{equation}
\noindent where $\Phi_r^2/2L_r=L_rI_r^2/2$ is the potential energy in the
CPW cavity,  and  the potential energy of the qubit is
\begin{equation}
U_i=E_{L,i}\left[\frac{1}{2}(\varphi_i-\varphi_{e,i})^2-\frac{L_i}{L_{J,i}}\cos{\varphi _i} \right]
\end{equation}
with $E_{L,i}=(\Phi_0/2\pi)^2/L_i$. The dimensionless flux,
$\varphi_{e,i}=2\pi \Phi_{i}/\Phi_0$, determines the profile of
the potential energy for the qubit. Using the Josephson relations
(to substitute $V_i$ with $\varphi_i$) and solving for the
equation of motion, after including the damping term, we
obtain\footnote{To speed up the calculations, the damping term for
the CPW cavity and the second qubit have been omitted}:

\begin{equation}
\begin{split}
&C_1\ddot {\varphi}_1 +C_{x}(\ddot{\varphi}_1-\ddot{\varphi}_r) =  -\frac{\partial U_1}{\partial \varphi_1}- \frac{1}{R_1}\dot{\varphi_1} \\
&C_r\ddot{\varphi}_r +C_{x}(\ddot{\varphi}_1-\ddot{\varphi}_r)
+C_{x}(\ddot{\varphi}_r-\ddot{\varphi}_2) = -\frac{\varphi_r}{L_r}
- \frac{\dot{\varphi_r}}{R_r}\\
&C_2\ddot{\varphi}_2  +C_{x}(\ddot{\varphi}_r-\ddot{\varphi}_2)  =
-\frac{\partial U_2}{\partial \varphi_2}-
\frac{1}{R_2}\dot{\varphi_2}
\end{split}\label{eq:eqofmotion}
\end{equation}

 where $\varphi_r=2\pi\Phi_r/\Phi_0$.
As can be seen from Eq.~\ref{eq:eqofmotion}, the treatment of the
qubits and the CPW cavity is fully classical. The initial
conditions for the solution of this system of differential
equations are described below. The second qubit and the oscillator
begin with zero kinetic energy
($\dot{\varphi_2}=\dot{\varphi_r}=0$) and have zero potential
energy; zero energy is defined at the bottom of the left well. To
understand the initial conditions for the first qubit it is useful
to recall the physics of the measurement. When the MP is applied,
the flux approaches the critical value (approximately
$0.95\varphi_{c}$) over a short period of time, so that the first
excited state tunnels out with unit probability. Once tunneled,
this qubit can be assumed to have zero kinetic energy
($\dot{\varphi_1}=0$), to have a phase value just to the right
side of the residual local maximum between the two wells
(Fig.~\ref{fig:QBpotential}(b)), giving it an initial potential
energy $\sim 0.2 \Delta U(\varphi_{e})$ below the local maximum
value\cite{kofmanprb2007}. In addition, we assume that the decay
rate in the right well is comparable to that  in the left well,
and the simulation is run for times $\sim 3T_1^1$, after which the
qubit phase has relaxed to rest. We have checked that small
variations in these assumptions do not meaningfully affect the
results of our simulations.
\begin{figure}
\includegraphics[width=\columnwidth,trim=0 3cm 0cm 2cm]{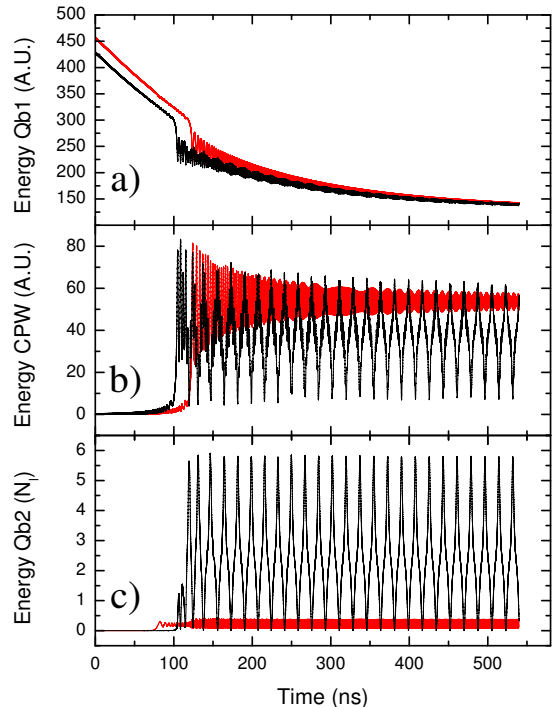}
\caption{(Color) Simulated energy  for  (a) the first qubit, (b) the CPW
cavity, (c) the second qubit. (Red):
$\varphi_1=0.8949\varphi_{c1}$ and $\varphi_2=0.893\varphi_{c2}$.
The first qubit decays exponentially up to $t\approx 123~ns$. At
this time the frequency of the oscillation in right well matches
the CPW cavity resonant frequency and the qubit transfers part of
its energy to the CPW cavity. The second qubit is resonating at a
different frequency and it is minimally exited by the incoming
microwave voltage. This corresponds to the red x of Fig.~2(b) (Black):
$\varphi_1=0.82\varphi_{c1}$ and
$\varphi_2=0.836\varphi_{c2}$.  In this case the first qubit
transfers part of its energy to the CPW cavity at $t\approx
103~ns$ because it starts at a lower energy in the deep well. At
this flux  the second qubit is in resonance with the cavity and it
is excited up to the sixth quantized level. This corresponds to the
white x of Fig.~2(b) 
\label{fig:singlelinecut}}
\end{figure}
From these initial conditions the phase of the first
qubit(classically) undergoes damped oscillations in the anharmonic
right well. Because of the anharmonicity of the potential, when
the amplitude of the oscillation is large, the frequency of the
oscillations is lower than the unmeasured qubit frequency. As the
system loses energy due to the damping, the oscillation frequency
increases as seen by the CPW cavity. When the crosstalk voltage
has a frequency close to the CPW cavity frequency, it can transfer
energy to the CPW cavity. If the second qubit's frequency matches
that of the CPW cavity then the cavity's excitation can be
transferred to the second qubit. In Fig.~\ref{fig:experiment}(b)
we plot, for the second qubit, the ratio ($N_l$) between the
maximum energy acquired and $\hbar \omega_p$, where $\omega_p$ is the plasma frequency of the qubit in the left well, as a function of the fluxes in the two qubits. The
crosstalk, measured as the maximum energy transferred to the
second qubit, is maximum at a flux $\varphi_2/\varphi_{c2}\sim
0.837$, where the second qubit's frequency is $\approx 8.97~GHz$,
determined by taking the Fast Fourier Transform of the
oscillations in energy over time (see Fig.~\ref{fig:singlelinecut}
(a-c)). Reversing the roles of the two qubits, we find that for
the first qubit the crosstalk is maximum at a flux $\sim
0.825\varphi_{c1}$, corresponding to an excitation frequency of
$\approx 8.84~GHz$ (Fig.~\ref{fig:experiment}(d)). These values
were determined for qubit 2 (qubit 1) by performing a Gaussian fit
of $N_l$ versus flux (or frequency) after averaging over the span
of flux (or frequency) values for qubit 1 (qubit 2). Notice that
the crosstalk transferred to qubit 2 (qubit 1) is flux independent
of qubit 1 (qubit 2) and substantial  only when the cavity
frequency matches the frequency of qubit 2 (qubit 1). The results
of the simulations are in  good agreement with the experimental
data.
To gain additional insight into the dynamics of the system, we
plot the time evolution of the energy for the qubits and the CPW
cavity (Fig.~\ref{fig:singlelinecut} (a-c)) for two different sets
of fluxes in the two qubits. At $\varphi_1=0.895\varphi_{c1}$ and
$\varphi_2=0.893\varphi_{c2}$ (red x in Fig.~\ref{fig:experiment}(b)) the first qubit decays exponentially for a time $t
\lesssim 123~ns$ (Fig.~\ref{fig:singlelinecut}
(a-c)-Red). At $t=123~ns$ there is a downward jump in the
energy of the first qubit while the energy of the CPW cavity
exhibits an upward jump. At this time, the frequency of
oscillation in the right well matches the CPW cavity resonant
frequency, so part of the qubit energy is transferred to the CPW
cavity. However, since   the second qubit is not on resonance with
the CPW cavity, it does not get significantly excited by the
microwave current passing through the capacitor $C_{x}$.

At $\varphi_1=0.82\varphi_{c1}$ and $\varphi_2=0.836\varphi_{c2}$
(white x in Fig.~\ref{fig:experiment}(b)),
 the dynamics of the
first qubit and the CPW cavity are essentially unchanged, except
that the CPW cavity frequency is matched at a different time
($t=103~ns$) because the first qubit starts at a lower energy in
the deep well (Fig.~\ref{fig:singlelinecut} (a-c)-Black). However, in this case, the second qubit is on
resonance with the CPW cavity and is therefore excited to an
energy $N_l\sim 6$.

The presence of the MP in the actual experiment does not change
the agreement between the experiment and the simulation. In fact,
the excitation arrives at the second qubit several nanoseconds
after the first qubit has been measured. Because of the finite
tunneling probability of the higher levels of the qubit, this
excitation can be sufficient to allow the qubit to tunnel through
the barrier.

We have shown experimental results for measurement crosstalk
between two metastable phase qubits coupled by a resonant cavity.
These results have been described classically and verified by a
full simulation of the system. We have confirmed that the resonant
cavity acts as a bandpass filter reducing the detrimental affects
of measurement crosstalk as long as the qubits are detuned from
the cavity resonance. In the future, it may be interesting to
investigate the transfer of measurement crosstalk through an
anharmonic resonant cavity whose frequency versus amplitude
characteristic may eliminate crosstalk completely. Ultimately,
future device architectures incorporating multiple metastable
phase qubits can benefit from the use of resonant cavities between
the qubits to prevent measurement crosstalk. This will improve the
fidelity and reduce errors in the measurement of entangled qubit
states\cite{AltomareX2009, KofmanPRB2008,AnsmannN2009}.

This work was financially supported by NIST. Contribution of the
U.S. government, not subject to copyright. M.A.S. was supported by the  
Academy of Finland, and by the ERC (grant No. FP7-240387).
%
%
%
%\bibliography{crosstalk}
%
%
%
%Merlin.mbs v4.21 2009-07-09.
%

\end{document}